\documentstyle[12pt,epsfig]{article}
\advance\textheight by \topskip
\begin{document}
\tolerance=100000
\thispagestyle{empty}
\setcounter{page}{0}

\def\cO #1{{\cal{O}}\left(#1\right)}
\newcommand{\be}{\begin{equation}}
\newcommand{\ee}{\end{equation}}
\newcommand{\br}{\begin{eqnarray}}
\newcommand{\er}{\end{eqnarray}}
\newcommand{\ba}{\begin{array}}
\newcommand{\ea}{\end{array}}
\newcommand{\bi}{\begin{itemize}}
\newcommand{\ei}{\end{itemize}}
\newcommand{\bn}{\begin{enumerate}}
\newcommand{\en}{\end{enumerate}}
\newcommand{\bc}{\begin{center}}
\newcommand{\ec}{\end{center}}
\newcommand{\ul}{\underline}
\newcommand{\ol}{\overline}
\newcommand{\ar}{\rightarrow}
\newcommand{\sm}{${\cal {SM}}$}
\newcommand{\as}{\alpha_s}
\newcommand{\aem}{\alpha_{em}}
\newcommand{\ycut}{y_{\mathrm{cut}}}
\newcommand{\susy}{{{SUSY}}}
\newcommand{\Dir}{\kern -6.4pt\Big{/}}
\newcommand{\Dirin}{\kern -10.4pt\Big{/}\kern 4.4pt}
\newcommand{\DDir}{\kern -10.6pt\Big{/}}
\newcommand{\DGir}{\kern -6.0pt\Big{/}}
\def\Ecm{\ifmmode{E_{\mathrm{cm}}}\else{$E_{\mathrm{cm}}$}\fi}
\def\gluino{\ifmmode{\mathaccent"7E g}\else{$\mathaccent"7E g$}\fi}
\def\photino{\ifmmode{\mathaccent"7E \gamma}\else{$\mathaccent"7E \gamma$}\fi}
\def\mgluino{\ifmmode{m_{\mathaccent"7E g}}
             \else{$m_{\mathaccent"7E g}$}\fi}
\def\taugluino{\ifmmode{\tau_{\mathaccent"7E g}}
             \else{$\tau_{\mathaccent"7E g}$}\fi}
\def\mphotino{\ifmmode{m_{\mathaccent"7E \gamma}}
             \else{$m_{\mathaccent"7E \gamma}$}\fi}
\def\ML{\ifmmode{{\mathaccent"7E M}_L}
             \else{${\mathaccent"7E M}_L$}\fi}
\def\MR{\ifmmode{{\mathaccent"7E M}_R}
             \else{${\mathaccent"7E M}_R$}\fi}

\def\Ord{\buildrel{\scriptscriptstyle <}\over{\scriptscriptstyle\sim}}
\def\OOrd{\buildrel{\scriptscriptstyle >}\over{\scriptscriptstyle\sim}}
\def\jp #1 #2 #3 {{\it J.~Phys.} {\bf#1} (#2) #3}
\def\pl #1 #2 #3 {{\it Phys.~Lett.} {\bf#1} (#2) #3}
\def\np #1 #2 #3 {{\it Nucl.~Phys.} {\bf#1} (#2) #3}
\def\zp #1 #2 #3 {{\it Z.~Phys.} {\bf#1} (#2) #3}
\def\pr #1 #2 #3 {{\it Phys.~Rev.} {\bf#1} (#2) #3}
\def\prep #1 #2 #3 {{\it Phys.~Rep.} {\bf#1} (#2) #3}
\def\prl #1 #2 #3 {{\it Phys.~Rev.~Lett.} {\bf#1} (#2) #3}
\def\mpl #1 #2 #3 {{\it Mod.~Phys.~Lett.} {\bf#1} (#2) #3}
\def\rmp #1 #2 #3 {{\it Rev. Mod. Phys.} {\bf#1} (#2) #3}
\def\sjnp #1 #2 #3 {{\it Sov. J. Nucl. Phys.} {\bf#1} (#2) #3}
\def\cpc #1 #2 #3 {{\it Comp. Phys. Comm.} {\bf#1} (#2) #3}
\def\xx #1 #2 #3 {{\bf#1}, (#2) #3}
\def\preprint{{\it preprint}}

\begin{flushright}
{DFTT 23/01}\\ 
{CERN-TH/2001-115}\\
\end{flushright}

\vspace*{\fill}

\begin{center}
{\Large \bf
Effects of jet algorithms from higher order QCD\\[0.25 cm]
in $W^\pm$ mass determinations at LEP2\footnote{Work supported
in part by the Ministero dell'Universit\`a e della Ricerca Scientifica and
by the European Union under contract HPRN-CT-2000-00149.
}}
\\[1.0 cm]
{\large E. Maina$^*$}\\[0.4 cm]
{\it Dipartimento di Fisica Teorica -- Universit\`a di Torino,}\\
{\it Istituto Nazionale di Fisica Nucleare -- Sezione di Torino,}\\
{\it Via Pietro Giuria 1, 10125 Torino, Italy.}\\[1.5cm]
{\large S. Moretti$^*$}\\[0.4 cm]
{\it CERN -- Theory Division}\\
{\it CH-1211 Geneva 23, Switzerland} 
\\[0.5cm]
\end{center}
\vspace*{\fill}

\begin{abstract}
{\noindent 
We analyse the impact of systematic effects due to the scale dependence of 
QCD corrections in combination with 
the use of different jet clustering algorithms 
in the measurement of the $W^\pm$ mass in the fully hadronic decay mode of
$W^+W^-$ pairs produced
at LEP2. We consider higher order contributions
induced by both virtual and real gluon radiation onto the electroweak
CC03 and CC11 channels through  
${\cal O}(\alpha_{s})$ at the parton level. We find that the 
associated uncertainties can be of order  
$100$ MeV, thus competitive with those possibly arising in the
non-perturbative regime and indeed above the current experimental
estimates.}
\end{abstract}
\vskip4.0cm
\noindent
$^*$ E-mails: maina@to.infn.it, stefano.moretti@cern.ch.
\vspace*{\fill}
\newpage

\section{Introduction}
\label{sec_intro}

Over the past years, LEP2 has been producing and studying $W^\pm$ bosons.
One of the main goals of such a collider was the determination of 
$M_{W^\pm}$ with a target accuracy of 40--50 MeV. This has apparently been
achieved. By combining their data
in all possible $W^\pm$ decay channels, the four LEP experiments quoted
the following result:  
\begin{equation}\label{MW}
M_{W^\pm}=80.427\pm0.046~{\mathrm{GeV}},
\end{equation}
that compares rather favourably with the estimates obtained
 at $p\bar p$ colliders \cite{LEPCEW}. This measurement is extremely important:
if combined with an improved determination of the top mass, $m_t$
(soon to be performed at the Tevatron, during Run 2), it can lead to a rather
stringent prediction of the Higgs mass, from a fit to
high precision electroweak (EW) data.

One of the experimental strategies adopted to measure the $W^\pm$ mass at LEP2
has been the kinematic reconstruction of the $W^\pm$ resonance through the
momenta of its decay products\footnote{An alternative method
is the so-called `threshold scan', wherein a value
for $M_{W^\pm}$ is fitted to the shape of the $e^+e^-\to W^+W^-$
total cross section for $\sqrt s$ in the vicinity of
$2 M_{W^\pm}$ (the result
quoted in eq.~(\ref{MW}) does also include measurements obtained in 
this way).}, 
e.g., in the fully hadronic channel:  
$e^+e^-\to W^+W^-\ar$ 4~jets. A `cleaner' measurement is certainly performed
in the semi-leptonic channel, i.e., $e^+e^-\to W^+W^-\ar$ 2 jets 
$\ell^\pm$ plus
an undetected neutrino (with $\ell=e,\mu$). However, the contemporaneous 
presence in this case of missing energy in the final state and of photon 
radiation in the initial state (ISR), loosens the kinematic constraints that 
can be applied in the $W^\pm$ mass reconstruction procedure. Besides, 
the fully  hadronic decay rate has a somewhat higher statistics than the
semi-leptonic one.  Therefore, although the 
event reconstruction is made harder in multi-jet final states 
by the larger number of tracks in the detector and by the usual 
uncertainties related to measuring jet energies and directions
(a task much less complicated in the case of leptons), 
 the $W^+W^-\to$ 4~jets mode represented an accurate
means of determining $M_{W^\pm}$ at LEP2 \cite{Budapest01}. In fact, as
shown in Ref.~\cite{Budapest01}, the separate results obtained
from the $W^+W^-\ar q\bar q' Q\bar Q'$ and $W^+W^-\to q\bar q'\ell\bar\nu_\ell$
channels are consistent, with a difference in mass which is very small:
\begin{equation}\label{deltaM}
\Delta M_{W^\pm}(q\bar q'Q\bar Q'-q\bar q'\ell\bar\nu_\ell)=+9\pm44~
{\mathrm MeV}.
\end{equation}
Nonetheless, systematics errors on $M_{W^\pm}$ are somewhat larger
in the $q\bar q'Q\bar Q'$ than in the $q\bar q'\ell\bar\nu_\ell$ channels:
see Table 2 of Ref.~\cite{Budapest01}.

Let us examine then more closely the kind of problems associated with
the $W^+W^-\ar$ 4~jets signature. One of the
issues is the problem of estimating theoretical biases due to the
relatively unknown `colour-reconnection' (CR) \cite{CR}
and `Bose-Einstein correlation' (BEC) \cite{bose} effects
(see \cite{Bryan} for a theoretical review and \cite{review} for
an experimental one). 
Things go as follows. In $e^+e^-\to W^+W^-\to4$~jets, one should expect
some interference effects between the two hadronic $W^\pm$ decays, 
simply because the decay products from the two different gauge bosons
can overlap considerably in space-time. In fact, at LEP2 energies, the
separation between the two $W^\pm$ decay vertices is $\sim0.1$ fm,
that is, much smaller than the typical hadronisation scale, $\sim1$ fm.
Hence, the two hadronic $W^\pm$ decays can no 
longer be considered as separate, since final-state interactions (CR) 
and/or identical-particle symmetrisation (BEC) can play a non-negligible
role, possibly leading to an apparent `shift' in the reconstructed $W^\pm$ mass
resonance \cite{Bryan,review}. 
Unfortunately, because of our current lack of understanding of 
non-perturbative QCD\footnote{In fact,
the perturbative effects of CR are expected to be small, because of
order  $\sim (C_F\alpha_s)^2/N_C^2\times\Gamma_{W^\pm}/M_{W^\pm}$
\cite{valtor} -- see also Refs.~\cite{CRparton1,CRparton2} -- so are 
partonic `Fermi-Dirac correlations' (as opposed to
hadronic BECs) \cite{WW6q}.}, such interference effects can only be
estimated theoretically in the context of different `models'. Most
of the latter can be constrained by looking at experimental
observables which are sensitive to either phenomenon. For example,
CR effects would be manifest in the central region, particularly
in the single-particle distributions at low momenta, as
produced in the fully hadronic versus the semi-leptonic channel, 
whereas BECs would lead to an increase in the correlation function
for $W^+W^-$, as compared to that for a single $W^\pm$ \cite{NewBryan}. 
While the jury is still out in the case of CR effects, there is an
increasing evidence that BEC effects are very small if at all present
\cite{WatsonMoriond01}.
In practice, the values that
the LEP experimental collaborations assign to the 
systematic errors on $M_{W^\pm}$
due to CR and BEC effects range from 30 to 66 MeV
and from 20 to 67 MeV, respectively (40 and 25 MeV are adopted
in the combined results) \cite{Budapest01}. 

Other problems in the fully hadronic decay channel of $W^+W^-$ pairs
are associated with the definition of `jets'. The problematic here
is twofold.
Firstly, because two identical decays take place in the same event, one has
the phenomenon of mis-pairing of jets. That is,  even
in the ideal case in which all tracks are correctly ascribed
to the parton from which they originate, one has to cope with the ambiguity
that it is in practice impossible to uniquely assign any pair among the four 
reconstructed jets to the parent $W^\pm$ on the sole basis of the 
event topology.  Of all possible combinations of di-jet systems, only one is 
correct. Thus, an
`intrinsic' background exists in $W^+W^-\ar$ 4~jets events, in terms of simple
combinatorics. Secondly, because of the large hadronic multiplicity,
one also has the phenomenon of mis-assignment of tracks.  In this case,
the ambiguity  stems from the fact that a track assigned to a 
jet, the latter eventually identified as a  parton
originating from one of the $W^\pm$'s, might have actually been produced 
in the fragmentation of another parton coming from the second $W^\mp$ decay.

Both these phenomenological aspects are clearly 
dependent upon the `jet clustering algorithm'  
(see Ref.~\cite{schemes} for a review), wherein the number 
of hadronic tracks is reduced one at a time by combining the two 
most (in some sense)
nearby ones (hereafter, we will quantify the `distance' between
two particles $i$ and $j$ by means of a variable denoted by 
$y_{ij}$). This (binary) 
joining procedure is stopped by means of a resolution 
parameter,  $\ycut$, and the final `clusters' yielding $y_{ij}$ values
all above $\ycut$ are called jets\footnote{Here
and in the following, the word `cluster' refers to hadrons or
calorimeter cells in the real experimental case, to partons in the
theoretical perturbative calculations, and also to intermediate jets
during the clustering procedure.}. 
In Ref.~\cite{schemes}, it was precisely this dependence that was
investigated, by using standard Monte Carlo (MC) simulation programs based on a
parton shower (PS) approach (see \cite{book}), such as HERWIG \cite{HERWIG}, 
JETSET/PYTHIA \cite{JETSET} and ARIADNE \cite{ARIADNE}. 
The results presented there did show a rather
dramatic effect in the reconstructed $M_{W^\pm}$ values, due to the
choice of the jet finder, and of its resolution parameter as well.
However, any shift on  $M_{W^\pm}$ of this sort can be estimated accurately,
as it is simply due to kinematic effects induced by the
jet clustering algorithm itself in reconstructing the quark
momenta starting from the PS (or after hadronisation). 
In practice, it can be treated as 
a well quantifiable
correction to be applied to the reconstructed $M_{W^\pm}$ value,
in order to reproduce the true one. Using the same MC programs, one can
also determine the typical size of the systematic errors due to the
hadronisation process, by comparing the outputs of the various programs.
Finally, background effects can be accounted for
by exploiting the numerous event generators available on the
market for $e^+e^-\to 4$ jets, both
in EW \cite{generators} and QCD \cite{HERWIG}--\cite{ARIADNE},
\cite{apacic} processes (see Ref.~\cite{spin} for a 
dedicated study of the impact of such QCD background
effects in $e^+e^-\to W^+W^-\to$ 4~jets).

Other systematic effects remain instead quite beyond control. These are 
intimately related to the way predictions are made within standard 
perturbation theory. That is, to the fact that only a finite number of terms of
a perturbative series are generally computed over all the available phase 
space. Or alternatively, that only over a restricted region of it,
all terms of a series can be summed to all orders. Whereas the availability
of the latter is in general more crucial to the estimation of a total cross
section, that of the former can be decisive for the study of more exclusive
observables.
Given the relative size of the EW and QCD coupling `constants'
at LEP2, it is clear that the dominant higher order effects will be due to   
the emission and/or absorption of gluons.

Several QCD effects entering $e^+e^-\to W^+W^-\to$ 4~jets
events have been studied so far.
For a start, it should be mentioned that 
the amplitude for $e^+e^-\ar W^+W^-\ar q\bar q' Q\bar Q'$ (the so-called
CC03 channel) is 
quite trivial to derive, in fact, more of a textbook example. It represents 
the lowest-order (LO) 
contribution to the $e^+e^-\to W^+W^-\ar$ $\mbox{4~jets}$ signal. 
Higher-order QCD
contributions involving gluons are, for example, the real ones 
(i.e., tree-level processes): $e^+e^-\ar W^+W^-\ar q\bar q' Q\bar Q' g$
and $e^+e^-\ar W^+W^-\ar q\bar q' Q\bar Q'  gg$ events, which have been 
calculated in Refs.~\cite{Brown} and \cite{CRparton1}, respectively, as well
as  $e^+e^-\ar W^+W^-\ar q\bar q' Q\bar Q' g^*$,
with the gluon splitting in a quark-antiquark pair, which was considered
in Ref.~\cite{WW6q} (see also \cite{short6}).
One-loop QCD corrections to $e^+e^-\ar W^+W^-\ar q\bar q' Q\bar Q'$
are also known to date \cite{loop}, and they have been interfered
with the LO amplitudes and eventually combined with the single real gluon 
emission contribution 
of Ref.~\cite{Brown} into the complete ${\cal O}(\alpha_s)$ result
\cite{loop}. In Ref.~\cite{CC11}, the full ${\cal O}(\alpha_s)$ 
corrections were computed for the case of the so-called CC11 
channel \cite{generators}, also including irreducible background
effects in addition to $e^+e^-\to W^+W^-$ production and decay. Finally,
two-loop effects due to the virtual exchange of two gluons between
the two quark pairs in hadronic $W^+W^-$ decays were estimated in 
Ref.~\cite{CRparton2} in the `soft limit' and found to be either small
(colour-singlet exchange) or large (colour-octet exchange) but symmetric
around $M_{W^\pm}$, hence unobservable in general.

We make use here of the calculations of Refs.~\cite{loop,CC11} in
order to assess
the size of the typical theoretical error due to the truncation
of the perturbative series at order  ${\alpha_s}$ and the systematic
effects that it introduces in observable quantities, primarily, in 
the `line-shape' of the
$W^\pm$ resonance, as determined by using different jet finders to select 
the hadronic sample. Our motivations to carry out such a study are
dictated by the following considerations.
For a start, NLO corrections
to both CC03 and CC11 have been found to be rather 
large in general \cite{loop,CC11}, with their actual size clearly    
depending upon the algorithm used. Furthermore, it is well known that
differential distributions are typically more sensitive (particularly
in presence of cuts over the phase space available to gluon emission)
to higher order effects than fully inclusive quantities, such as total
cross sections, where virtual and real contributions tend to cancel to 
a larger extent.
Besides, in the case of the $W^\pm$-mass line-shape, one would expect
the distortion effects to be mainly induced by relatively hard and
non-collinear gluons, which should be better modelled by an exact NLO
calculation than by the PS models exploited in Ref.~\cite{schemes}.

\section{Results}
\label{sec_results}

All QCD predictions have an intrinsic dependence on an arbitrary 
scale, hereafter denoted by $\mu$, entering at any order in $\as$. 
This scale is not fixed a priori. On the one hand, 
although the structure of the QCD perturbative expansion does not
prescribe which value should be adopted for  $\mu$, an
obvious requirement is that it should be of the order of the energy
scale involved in the problem: i.e., the CM energy $\sqrt s$ (see
Ref.~\cite{subtraction} for detailed discussions). 
On the other hand, the physical scales of
gluon emissions that actually give rise to multi-jet configurations
are to be found down to the energy scale 
${\sqrt{{y}_{\mathrm{cut}}s}}$. 
In practice, one should avoid building up large logarithmic
terms related to the (unphysical) `mismatch' between the process scale 
$\mu\approx \sqrt s$ and the emission
 scale ${\sqrt{{y}_{\mathrm{cut}}s}}$ and
it is well known that it may be necessary to adopt a different scale for
each observable in order to best describe experimental data taken at fixed    
$\sqrt s$ \cite{shapevar}.
It is precisely the $\mu$-dependence of the
truncated perturbative series that is treated very differently by
each jet-clustering algorithm \cite{schemes,cambridge,BKSS}
and the corresponding effects on observable quantities are what we aim to study. 
In our calculation, QCD effects appear through ${\cal O}(\alpha_s)$
only, so that the $\mu$-dependence is merely the one affecting
the strong coupling constant at lowest order. Whereas its impact is
trivial to assess (and account for) in the case of total inclusive rates,
this is no longer true for differential quantities (such as
mass distributions), because of the different kinematics of 
lowest and higher order contributions, respectively.
Notice, however,  that a problem arises when studying the scale 
dependence of $\as$ results for
algorithms based on different measures, as for the same $y_{\mathrm{cut}}$ the
total cross section at NLO can be significantly different.  A more
consistent procedure was outlined in Ref.~\cite{BKSS}: that is, to
compare the NLO scale dependence of the various schemes not at the
same $y_{\mathrm{cut}}$ value, rather at the same LO rate. This is
our approach.

In order to make more manifest the effects of the interplay 
between the ${\cal O}(\as) $ corrections to $W^+W^-\to 4$~partons events and 
the jet clustering schemes tested,
we have not considered here Coulomb corrections to CC03 \cite{coulomb}
(their relevance is anyway modest beyond the $W^+W^-$ threshold). 
We also have neglected
the implementation of the mentioned BEC and 
CR phenomena, as these mainly arise in the non-perturbative domain.
Similarly,  hadronisation and detector effects were not investigated, nor
those due to ISR. We refer the reader to Refs.~\cite{schemes}
and \cite{CC11}, respectively, where their impact was studied in detail.

As centre-of-mass (CM) energy representative of LEP2 we have used the value
 $\sqrt s=175$  GeV. As for the 
parameters of the theory,
we have adopted (in the fixed-width approach)
                $M_{Z^0}=91.189$ GeV, $\Gamma_{Z^0}=2.497$ GeV,
              $M_{W^\pm}=80.430$ GeV, $\Gamma_{W^\pm}=2.087$ GeV, 
$\sin^2 \theta_W=0.231$, $\alpha_{{em}}= 1/128.07$ and the one-loop expression
for $\alpha_{s}$ (for consistency), with 
$\Lambda_{\tiny{\mbox{QCD}}}^{N_F=4}=0.283$ GeV
(yielding $\alpha_s(\sqrt s)=0.123$)\footnote{The one-loop
values of $\alpha_s$ adopted here are consistent with the two-loop
one extracted from experimental fits to shape variables at the $Z^0$ peak,
$\alpha_s(M_{Z^0})=0.116$, for the same choice of 
$\Lambda_{\tiny{\mbox{QCD}}}^{N_F=4}$.}. Furthermore, 
we have kept all quarks massless as a default, 
in order to speed up the numerical evaluations. 
Electron and positron have mass zero too, so has the neutrino.
Also,  we have neglected Cabibbo-Kobayashi-Maskawa (CKM) mixing terms
(i.e., we have taken the CKM rotation matrix to be diagonal).

\begin{figure}[!t]
\begin{center}
~{\epsfig{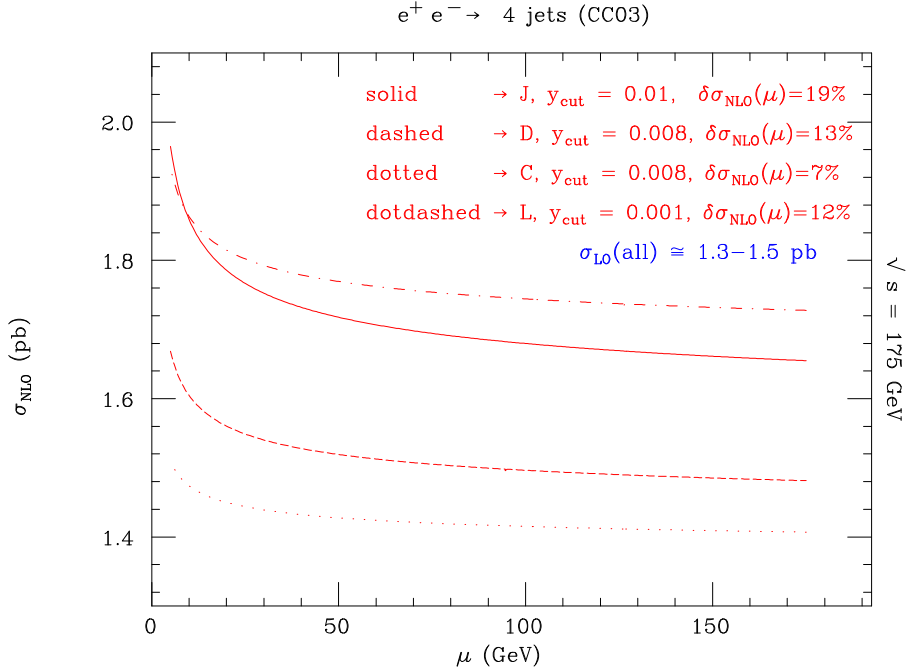}}
\end{center}
\begin{center}
\caption{NLO cross sections for $e^+e^-\to W^+W^-\to u\bar d s\bar c$ (CC03),
as a function of the QCD scale $\mu$,
as given by four different jet clustering algorithms, at $\sqrt s=175$  GeV.
(The values of $y_{\mathrm{cut}}$ are chosen such that the LO rates
are approximately equal for all schemes.)
Total hadronic rates are obtained by multiplying those above times four.}
\label{fig:CC03_NLO}
\end{center}
\end{figure}

As jet clustering schemes\footnote{We 
acknowledge here the well admitted abuse in referring to the various
jet `finders'
both as algorithms and as schemes, since the last term was originally
intended to identify the composition law of four-momenta when pairing two
clusters: in our case,
the so-called E-scheme, i.e., $p^\mu_{ij}=p^\mu_{i}+p^\mu_{j}$
(other choices have negligible impact on our conclusions).}, 
we have used a selection of
the binary ones, in which only two objects are clustered together at any step. 
 These are the following.
The JADE (J) one \cite{jade}, which uses as a measure of separation 
(or `metric') the quantity
\begin{equation}\label{J}
y_{ij}^J = {{2 E_i E_j (1-\cos\theta_{ij})}\over{s}}.
\end{equation}
The Durham (D) \cite{durham} and the Cambridge (C)
\cite{cambridge}  ones, both using 
\begin{equation}\label{D_C}
y_{ij}^D \equiv y_{ij}^C = {{2\min (E^2_i, E^2_j)(1-\cos\theta_{ij})}\over{s}}.
\end{equation}
(The Cambridge algorithm in fact only modifies 
 the clustering procedure of the Durham jet finder.)
We also have adopted the LUCLUS or LUND (L) jet finder \cite{luclus}, 
for which one has
\begin{equation}\label{L}
y_{ij}^L = \frac{2 |p_i|^2 |p_j|^2 (1 - \cos\theta_{ij})}%
{(|p_i| + |p_j|)^2 s},
\end{equation}
however, with the same clustering procedure of the Cambridge scheme
and without `preclustering' and `reassignment' (see Ref.~\cite{luclus}),
i.e., as done in Ref.~\cite{schemes} (where it was labelled as CL).
In eqs.~(\ref{J})--(\ref{L}),
$E_i(|p_i|)$ and $E_j(|p_j|)$ are the energies(moduli of the 
tree-momenta) and $\theta_{ij}$ the angular separation
of any pair $ij$ of particles in the final state, to be 
compared against the resolution parameter $y_{\mathrm{cut}}$. 
The choice of these particular schemes has a simple motivation.
The D, C and L ones are different versions of `transverse-momentum'
based algorithms, whereas the J one uses an `invariant-mass' measure
(the numerator of eq.~(\ref{J}) coincides with the invariant mass of
the partons $ij$, when the latter are massless, as is the case here).
In fact, these two categories are those that have so far been employed 
most in phenomenological studies of jet physics
in electron-positron collisions, with the former
gradually overshadowing the latter, thanks to their reduced
scale dependence in higher order QCD, e.g., in the case of the
${\cal{O}}(\as^2)$ three- \cite{schemes,BKSS,mb}
and ${\cal{O}}(\as^3)$ four-jet rates 
\cite{as4}, and to smaller hadronisation effects in the same contexts
\cite{schemes,BKSS}.

Our multi-jet sample is selected at the parton level, by requiring 
a final state with {\sl at least} four resolved objects (i.e., with
all their $y_{ij}$'s above  a given $y_{\mathrm{cut}}$). When five
objects survive, the two yielding the smallest $y_{ij}$ value
(according to the metric used)
are joined together, so to always produce a four-particle final state.
In doing so, we conform to typical experimental approaches: 
see, e.g., Ref.~\cite{example}.  The impact of a different treatment
of five-jet contributions was assessed in Ref.~\cite{loop}.

Fig.~\ref{fig:CC03_NLO} illustrates the dependence of the CC03 NLO rates
upon the unknown $\mu$ scale, for our four default
jet clustering algorithms, for representative choices of 
$y_{\mathrm{cut}}$ such
that the LO rates in the various schemes are approximately equal.
(In fact, differences at LO are typically within 15\%;
the actual numbers being: 
$\sigma_{LO}=1.38(1.32)[1.32]\{1.54\}$ pb for J(D)[C]\{L\}.)
 The variation of
the NLO rates with $\mu$, for values of the latter ranging
between $5$ GeV and $\sqrt s$\footnote{Note that by restraining 
$\mu$ to values higher than the hadronisation scale $Q_0$, 
which is of order 1 GeV, a perturbative analysis
is in principle
always justified. However, too low a value of $\mu$ would
imply a very large $\alpha_s$, in turn rendering the fixed order predictions
unreliable. As a compromise, we will be considering $\mu$-values 
well above $Q_0$ in the reminder of our study (say, 35 GeV and above).
For scale choices in the interval 35 GeV $<\mu<$ 175 GeV, the
strong coupling constant varies over the following ranges: 
0.162(0.134) $<\as<$ 0.123(0.105) at one-(two-)loop level.}, 
denoted by $\delta\sigma_{\mathrm{NLO}}(\mu)$, depends upon the
jet algorithm, varying significantly,
between 7\% (C scheme) and 19\% (J scheme).
The $K$-factors, for $\mu=\sqrt s$, are also very different, from one
algorithm to another, again with the minimum corresponding to the
C scheme ($K=1.07$) and the maximum to the J one ($K=1.20$). 
All these values
are however lower limits. In fact, as $\mu$ is decreased $\as$ increases, 
hence the relative size of the NLO effects grows larger too, in each case. 

\begin{figure}[!t]
\begin{center}
~{\epsfig{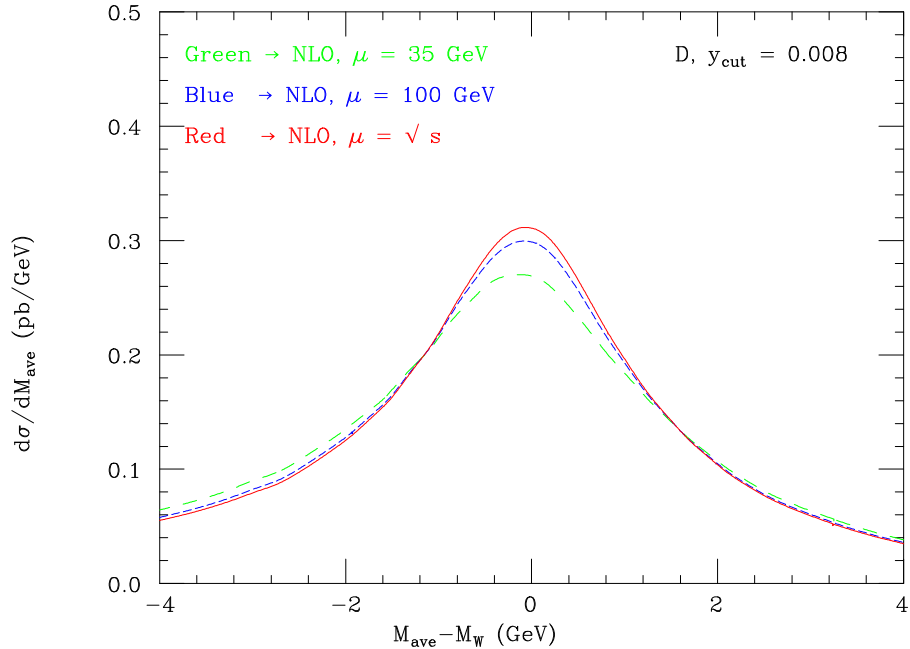}}
\end{center}
\begin{center}
\caption{The zero-th (LO) and first order (NLO--LO) $\alpha_s$ components
of the NLO differential distribution in the `average' mass
(as defined in the text) for $e^+e^-\to W^+W^-\to u\bar d s\bar c$ (CC03),
the latter with QCD scale $\mu=\sqrt s$,
as given by four different jet clustering algorithms, at $\sqrt s=175$  GeV.
(The values of $y_{\mathrm{cut}}$ are chosen such that the LO rates
are approximately equal for all schemes.) Notice that the dashed and
dotted blue-lines are visually indistinguishable.
Total hadronic rates are obtained by multiplying those above times four.}
\label{fig:CC03_finefunction}
\end{center}
\end{figure}

\begin{figure}[!t]
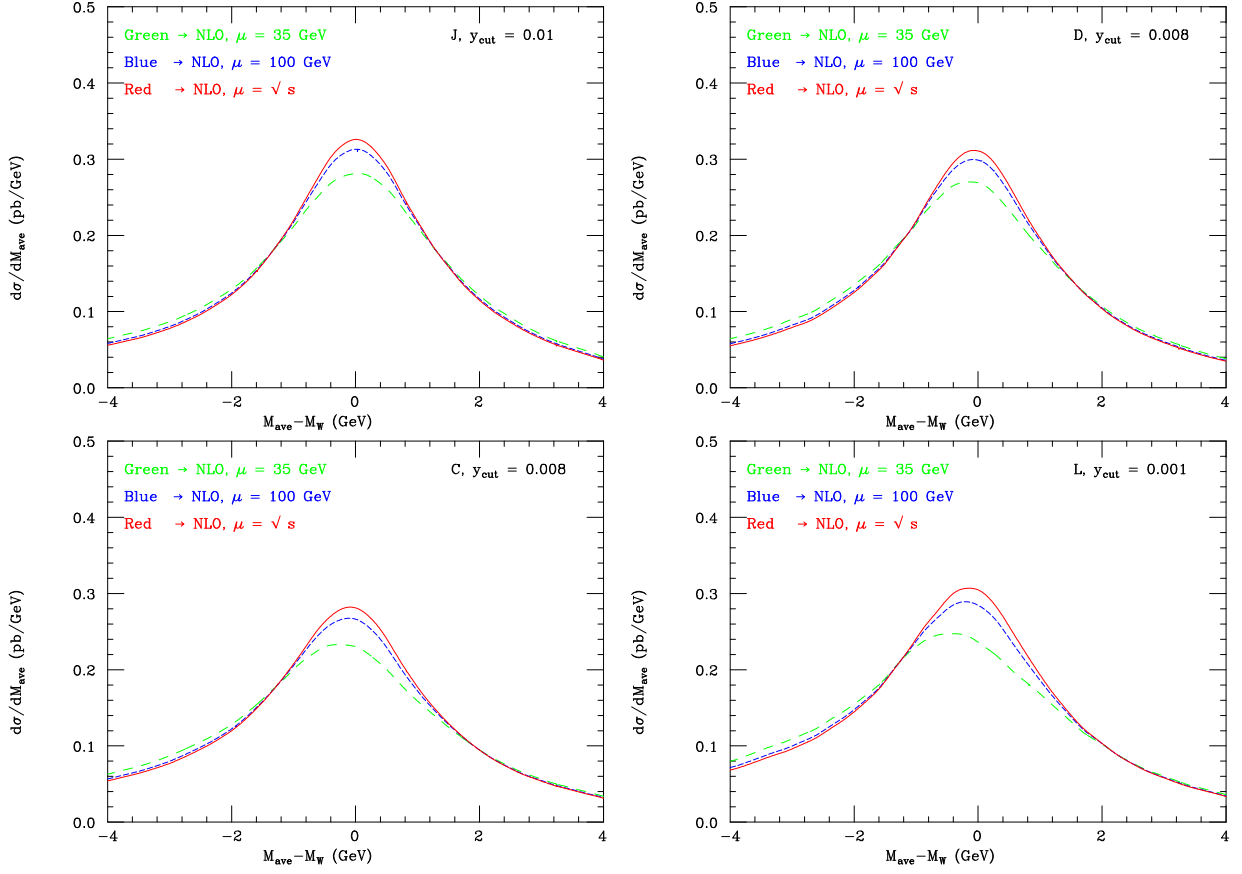

\begin{center}
~{\epsfig{file=StefanoMoretti_fig3fonts.ps,height=3.5cm,angle=90}}\\[0.25truecm]
~{\epsfig{file=StefanoMoretti_fig3a.ps,height=8cm,angle=90}}
~{\epsfig{file=StefanoMoretti_fig3b.ps,height=8cm,angle=90}}\\
~{\epsfig{file=StefanoMoretti_fig3c.ps,height=8cm,angle=90}}
~{\epsfig{file=StefanoMoretti_fig3d.ps,height=8cm,angle=90}}\\
\end{center}
\begin{center}
\caption{NLO differential distribution in the `average' mass
(as defined in the text) for $e^+e^-\to W^+W^-\to u\bar d s\bar c$ (CC03),
for three choices of the QCD scale $\mu$,
as given by four different jet clustering algorithms, at $\sqrt s=175$  GeV.
(The values of $y_{\mathrm{cut}}$ are chosen such that the LO rates
are approximately equal for all schemes.)
Total hadronic rates are obtained by multiplying those above times four.}
\label{fig:CC03_mu}
\end{center}
\end{figure}

As an estimator for the $W$ mass we use the `average' mass, $M_{\mathrm{ave}}$,
defined as follows.
Out of the three possible combinations of pairs of jet-jet systems, we
choose the one for which the two reconstructed $W^\pm$ masses, $M_{R_1}$
and $M_{R_2}$, minimise 
\begin{equation}\label{minimise}
\Delta M=|M_{R_1}-M_{W^\pm}|+|M_{R_2}-M_{W^\pm}|
\end{equation}
 and then define
\begin{equation}\label{average}
M_{\mathrm{ave}}=\frac{1}{2}(M_{R_1}+M_{R_2}).
\end{equation}
This variable has been extensively used since, at tree level, the difference between
$M_{\mathrm{ave}}$ and the average between the two $W^\pm$ masses
that one would reconstruct if the quarks could always be paired correctly
is very small\footnote{One could consider more
sophisticated approaches, but this is beyond the scope of this
paper.}.

The NLO differential spectra in $M_{\mathrm{ave}}$
show a shift towards low mass values with respect to 
the LO case which depends upon the jet algorithm being used
and its $y_{\rm{cut}}$ value. 
(The generation of this low mass tail at NLO was already observed 
and discussed in Ref.~\cite{loop}, where its consequences for a determination
of $M_{W^\pm}$ were described in details.)

The NLO distribution in a generic mass $M$ is made up by two terms:
\begin{equation}
\label{eq:AB}
d\sigma_{\mathrm{NLO}}/dM=A(M)+\alpha_s B(M),
\end{equation}
one proportional to the zero-th power of $\alpha_s$ (denoted by
$A$: the Born term) and another to the first power (denoted by
$B$: the first order correction). They are plotted separately
in Fig.~\ref{fig:CC03_finefunction} for the mass difference
$M=M_{\mathrm{ave}}-M_{W^\pm}$, labelled as `LO' and
`NLO--LO', respectively. Numerical values of the two
functions for some selected `average' masses are
given in Tab.~\ref{tab:CC03_AB} (where $M=M_{\mathrm{ave}}$), both 
normalised here to the total Born cross section (i.e., the integral over
$M_{\mathrm{ave}}$ of the LO curves in Fig.~\ref{fig:CC03_finefunction}).
The scale of the strong coupling constant is still set to $\mu=\sqrt s$.

However, we are concerned here with the fact that
the actual shape of the $M_{\mathrm{ave}}-M_{W^\pm}$ distribution
at NLO depends upon $\mu$ (this was set
to $\sqrt s$ as default in \cite{loop}), through the choice of both
the jet finder and its resolution parameter. Ultimately then,
so will do the value of $M_{W^\pm}$ extracted from that
distribution. To study this effect, we
plot in  Fig.~\ref{fig:CC03_mu} the differential distribution of the 
quantity defined in eqs.~(\ref{minimise})--(\ref{average}), now 
for three different choices of $\mu$, e.g.,
$35,100$ GeV and $\sqrt s$, for our default choice of
jet algorithms and $y_{\mathrm{cut}}$'s. There exists a visible
variation with $\mu$; besides, the previously
observed dependence on the choice of the algorithm and
$y_{\rm{cut}}$ persists at different $\mu$'s.
  
\begin{table}[htbp]
\begin{center}
\begin{tabular}{|c||c|c||c|c|}
\hline
\rule[0cm]{0cm}{0cm}
 &
\multicolumn{4}{c|}{$d\sigma_{\mathrm{NLO}}/dM_{\mathrm{ave}}/\sigma_{\mathrm{LO}}$ (1/GeV)} \\ 
\hline
\hline
\rule[0cm]{0cm}{0cm}& \multicolumn{2}{c||}{J \ $y_{\mathrm{cut}}=0.01$} & \multicolumn{2}{c|}{D \ $y_{\mathrm{cut}}=0.008$} 
\\ \hline
$M_{\mathrm{ave}}$ (GeV)
& $A(M_{\mathrm{ave}})$& $\alpha_s B(M_{\mathrm{ave}})$ & $A(M_{\mathrm{ave}})$& $\alpha_s B(M_{\mathrm{ave}})$\\ \hline
  72.43 &   0.00484 &   0.01473 &   0.00420 &   0.01303 \\ 
  73.43 &   0.00614 &   0.01663 &   0.00584 &   0.01507 \\ 
  74.43 &   0.00922 &   0.01688 &   0.00886 &   0.01610 \\ 
  75.43 &   0.01346 &   0.01808 &   0.01382 &   0.01936 \\ 
  76.43 &   0.02115 &   0.01973 &   0.02075 &   0.02072 \\ 
  77.43 &   0.03743 &   0.01993 &   0.03521 &   0.02712 \\ 
  78.43 &   0.07545 &   0.01489 &   0.07352 &   0.02412 \\ 
  79.43 &   0.18597 &  --0.02111 &   0.18431 &  --0.00957 \\ 
  80.43 &   0.33701 &  --0.10389 &   0.34214 &  --0.10448 \\ 
  81.43 &   0.16682 &  --0.01837 &   0.16996 &  --0.02670 \\ 
  82.43 &   0.06588 &   0.01480 &   0.06718 &   0.00691 \\ 
  83.43 &   0.03155 &   0.01351 &   0.03359 &   0.01200 \\ 
  84.43 &   0.01699 &   0.00697 &   0.01822 &   0.00743 \\ 
\hline
\hline
\rule[0cm]{0cm}{0cm}
&\multicolumn{2}{c||}{C \ $y_{\mathrm{cut}}=0.008$} &\multicolumn{2}{c|}{L \ $y_{\mathrm{cut}}=0.001$} \\ \hline
$M_{\mathrm{ave}}$ (GeV)
& $A(M_{\mathrm{ave}})$& $\alpha_s B(M_{\mathrm{ave}})$ & $A(M_{\mathrm{ave}})$& $\alpha_s B(M_{\mathrm{ave}})$\\ \hline
  72.43 &   0.00418 &   0.01348  &   0.00468 &   0.01696 \\ 
  73.43 &   0.00582 &   0.01586  &   0.00660 &   0.01838 \\ 
  74.43 &   0.00889 &   0.01651  &   0.00909 &   0.01895 \\ 
  75.43 &   0.01379 &   0.01853  &   0.01428 &   0.02218 \\ 
  76.43 &   0.02079 &   0.02184  &   0.02122 &   0.02423 \\ 
  77.43 &   0.03517 &   0.02518  &   0.03694 &   0.02446 \\ 
  78.43 &   0.07351 &   0.02265  &   0.07447 &   0.02048 \\ 
  79.43 &   0.18431 &  --0.01907  &   0.18530 &  --0.02824 \\ 
  80.43 &   0.34205 &  --0.12461  &   0.34140 &  --0.14456 \\ 
  81.43 &   0.16998 &  --0.03907 &   0.16612 &  --0.04661 \\ 
  82.43 &   0.06717 &   0.00227  &   0.06619 &  --0.00090 \\ 
  83.43 &   0.03360 &   0.00716  &   0.03181 &   0.00219 \\ 
  84.43 &   0.01822 &   0.00433  &   0.01677 &   0.00581 \\ 
\hline

\end{tabular}
\caption{The $A(M_{\mathrm{ave}})$ and $\alpha_s B(M_{\mathrm{ave}})$
components of the NLO cross section normalised to the Born rate,
see eq.~(\ref{eq:AB}), for representative values of $M_{\mathrm{ave}}$,
as obtained by our default jet clustering algorithms and
resolutions, with $\alpha_s = 0.123$ evaluated through one-loop
order at the scale $\mu=\sqrt s=175$ GeV.
Recall that the input value for the $W^\pm$ mass is $80.430$ GeV.}
\label{tab:CC03_AB}
\end{center}
\end{table}

\begin{figure}[!t]
\begin{center}
~{\epsfig{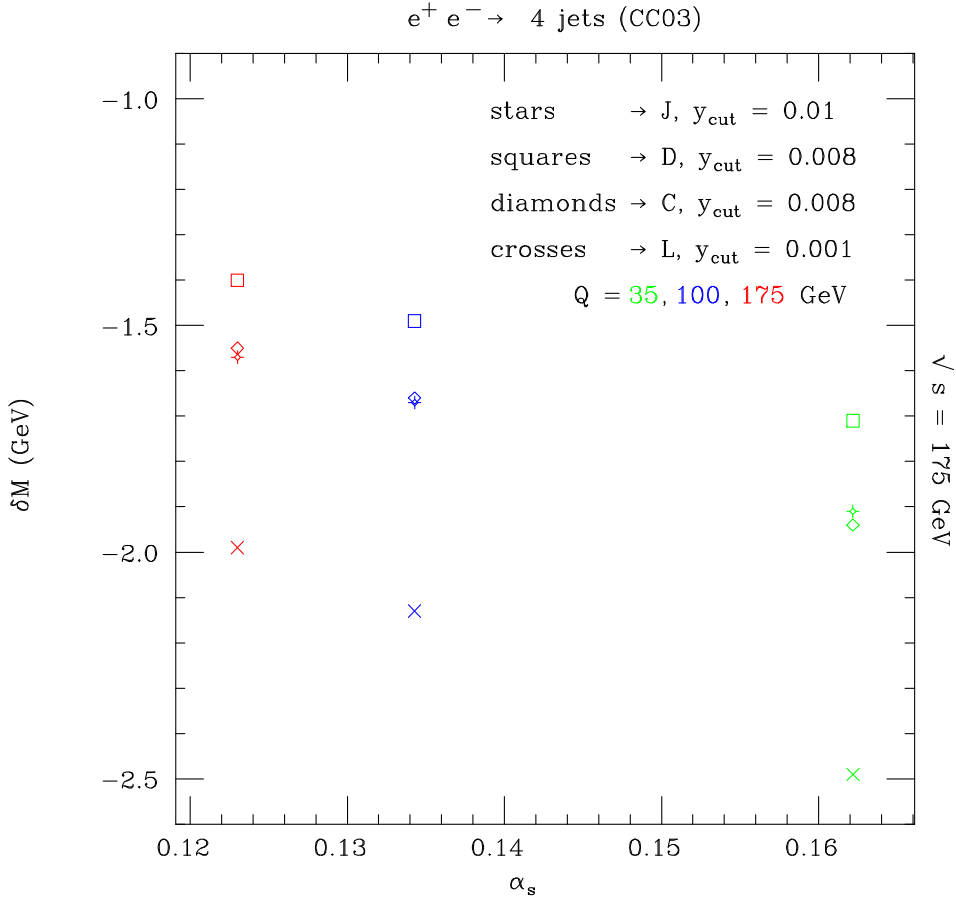}}
\end{center}
\begin{center}
\caption{The NLO mass difference
$\delta M = <M_{\mathrm{ave}}-M_{W^\pm}>$  
as a function of $\alpha_s$ calculated at
one-loop in $e^+e^-\to W^+W^-\to u\bar d s\bar c$ (CC03),
for three choices of the QCD scale $\mu$,
as given by four different jet clustering algorithms, at $\sqrt s=175$  GeV.}
\label{fig:newshift}
\end{center}
\end{figure}

In order to quantify the impact of the $\mu$-dependence of 
the $W^+W^-\to4$ jet rates on the  ${W^\pm}$ mass, we
have collected in Tab.~\ref{tab:CC03_std} the values obtained 
for the mean deviation from $M_{W^\pm}$
of the reconstructed `average' mass  $M_{\mathrm{ave}}$, as
predicted by our default jet clustering algorithms and separations.
Both at LO and NLO, the difference $<{{M}}_{\mathrm{ave}}-M_{W^\pm}>$
is negative, as already observed in \cite{loop}. Besides, by comparing
the higher order predictions obtained with $\mu$ varying from 35 to 175 GeV, 
one may notice that systematic uncertainties on $M_{W^\pm}$
could turn out be very large in the end, since
$<{{M}}_{\mathrm{ave}}-M_{W^\pm}>$ can be as large as 500 MeV
(in the L scheme).
The same data are plotted as a function of $\alpha_s$ in 
Fig.~\ref{fig:newshift}. For all schemes the $\alpha_s$ dependence is
essentially linear and the shift for different ranges in $\alpha_s$
is readily evaluated.

\begin{table}[htbp]
\begin{center}
\begin{tabular}{|c||c|c|c|}
\hline
\rule[0cm]{0cm}{0cm}
& \multicolumn{3}{c|}{$<{{M}}_{\mathrm{ave}}-M_{W^\pm}>$
(GeV)} \\ \hline
\rule[0cm]{0cm}{0cm}
Algorithm & {$\mu=\sqrt s/5=35$ GeV} &
            {$\mu=\sqrt s=100 $ GeV} &
            {$\mu=\sqrt s=175 $ GeV} \\ \hline
\rule[0cm]{0cm}{0cm}
 J    & $-0.34$ & $-0.34$ & $-0.34$ \\ 
      & $-1.91$ & $-1.67$ & $-1.57$ \\ \hline
~D    & $-0.28$ & $-0.28$ & $-0.28$ \\ 
      & $-1.71$ & $-1.49$ & $-1.40$ \\ \hline
~C    & $-0.28$ & $-0.28$ & $-0.28$ \\ 
      & $-1.94$ & $-1.66$ & $-1.55$ \\ \hline
~L    & $-0.35$ & $-0.35$ & $-0.35$ \\ 
      & $-2.49$ & $-2.13$ & $-1.99$ \\ \hline
\end{tabular}
\caption{Mean difference between $M_{\mathrm{ave}}$ and $M_{W^\pm}$
as obtained from (some of) the spectra in 
Figs.~\ref{fig:CC03_finefunction}--\ref{fig:CC03_mu}. First line
is for LO results (these do not depend upon
$\mu$), second line is for the NLO ones. 
Recall that the input value for the $W^\pm$ mass is $80.430$ GeV.}
\label{tab:CC03_std}
\end{center}
\end{table}

In order to estimate more realistically the systematic uncertainty on the
determination of the $W^\pm$ mass induced by the unknown 
scale $\mu$, we perform a {\tt MINUIT} \cite{minuit}
fit on the LO and NLO distributions, with a fitting function of the form
\begin{equation}\label{fitf}
f(m)=c_1\frac{c_2^2 c_3^2}{(m^2-c_2^2)^2+c_2^2 c_3^2}+g(m)
\end{equation}
where the term $g(m)$ is meant to simulate a smooth background
due to mis-assigned jets induced by the clustering algorithm.
For the latter, we adopt two possible choices
\begin{eqnarray}\label{fitg}
g(m) = \left \{ \begin{array}{c}  
                                  c_4+c_5~(m-c_2)+c_6~(m-c_2)^2, \\[3mm]
                                  c_4\frac{1}{1+{\mathrm{exp}}((m-c_5)/c_6)},
\end{array} \right. 
\end{eqnarray}
that is, a three-term polynomial and a smeared
step function (motivated by the kinematical-limit shoulder at large masses
and on the same footing as in Ref.~\cite{schemes}).
Notice that in eq.~(\ref{fitf}) we have implicitly assumed a Breit-Wigner 
shape characterised by a peak height $c_1$, a position $c_2$ and a width $c_3$,
corresponding to the normalisation 
$h$\footnote{This is related to the input
normalisation for {\tt MINUIT} and is irrelevant to our
purposes.}, 
$M_{W^\pm}$ and $\Gamma_{W^\pm}$, 
respectively, of the  distributions in 
Figs.~\ref{fig:CC03_finefunction}--\ref{fig:CC03_mu}. 
To first approximation, the difference between the values
of the coefficient $c_2$ as obtained from fitting the above curves 
 is then a measure of the typical
size of the systematic error that we are investigating.
Obviously, more sophisticated fitting procedures could be adopted, possibly
yielding different results for $M_{W^\pm}$.
However, it should be clear from a close inspection of the plots in 
Fig.~\ref{fig:CC03_mu} and
the estimates in Tab.~\ref{tab:CC03_std} that varying $\mu$ over any
reasonable interval would result in mass shifts comparable to
or larger than the uncertainty reported in (\ref{MW}).

\begin{table}[htbp]
\begin{center}
\begin{tabular}{|l|r|r|r|r|r|r|}
\hline
\rule[0cm]{0cm}{0cm}
Algorithm & \multicolumn{3}{c|}{$\mu=\sqrt s/5=35$ GeV} &
\multicolumn{3}{c|}{$\mu=\sqrt s=175$ GeV} \\ \hline
\rule[0cm]{0cm}{0cm}
   & \multicolumn{1}{c|}{$h$} & $M_{W^\pm}$ & $\Gamma_{W^\pm}$ &
     \multicolumn{1}{c|}{$h$} & $M_{W^\pm}$ & $\Gamma_{W^\pm}$ \\
   &  & (GeV) & (GeV) &   & (GeV) & (GeV)   \\ \hline
\rule[0cm]{0cm}{0cm}
 & \multicolumn{6}{c|}{Polynomial background} \\ \hline
\rule[0cm]{0cm}{0cm}
 J   
      & 517.757 & 80.464 & 3.237 & 575.412 & 80.454 & 2.828\\ \hline
~D   
      & 480.344 & 80.298 & 3.163 & 539.973 & 80.343 & 2.788\\ \hline
~C   
      & 422.196 & 80.211 & 3.450 & 486.111 & 80.292 & 2.930\\ \hline
~L   
      & 453.062 & 80.085 & 3.801 & 529.413 & 80.217 & 3.142\\ \hline
\rule[0cm]{0cm}{0cm}
 & \multicolumn{6}{c|}{Smeared step function background} \\ \hline
\rule[0cm]{0cm}{0cm}
 J   
      & 516.169 & 80.451 & 3.213 & 601.842 & 80.472 & 3.008\\ \hline
~D   
      & 521.475 & 80.286 & 3.450 & 549.312 & 80.332 & 2.890\\ \hline
~C   
      & 454.114 & 80.226 & 3.619 & 517.671 & 80.320 & 3.163\\ \hline
~L   
      & 456.933 & 80.154 & 3.791 & 515.275 & 80.221 & 3.127\\ \hline
\end{tabular}
\caption{Fits to (some of) the $M_{\mathrm{ave}}$ spectra at NLO in 
Figs.~\ref{fig:CC03_finefunction}--\ref{fig:CC03_mu}. First three columns are 
for $\mu=\sqrt s/5$, 
last three are for $\mu=\sqrt s$,
with $\sqrt s=175$ GeV. A Breit-Wigner shape is always
assumed, supplemented by a three-term polynomial (upper section)
or a smeared step function (lower section) to emulate
the intrinsic  background. 
Here, we have fitted the 
$M_{\mathrm{ave}}$  distributions over the mass interval 75 to 85 GeV.
Recall that the input value for the $W^\pm$ mass is $80.430$ GeV.}
\label{tab:CC03_fit}
\end{center}
\end{table}

Tab.~\ref{tab:CC03_fit} reproduces the results of one of our 
fits. Whereas the values of some of the parameters (such as the height $h$ and
the width $\Gamma_{W^\pm}$ of the distributions) depend sensibly on the 
choice of the mass interval used for the fit and/or the form of the background,
the values extracted for $M_{W^\pm}$ at NLO
are remarkably more stable\footnote{For reason of space,
we do not reproduce here the values of the coefficients
$c_4, c_5$ and $c_6$ which characterise the background.}. 
The parameters in the table should be taken as
representative of the qualitative features of all fits that we 
performed.  From there, one
notices a strong dependence of the fitted $M_{W^\pm}$ value upon the 
jet scheme (for a `fixed' $\mu$), with variations of up to 
almost 380 MeV (between the J and  L schemes,
when $\mu=35$ GeV and assuming a polynomial background). 

However, all effects discussed above are well understood,
since they are merely of kinematical origin (the different handling of  
gluon radiation by the various jet clustering algorithms).
As already stressed repeatedly, it is
the systematics associated with the choice of $\mu$ that is beyond
theoretical control. It turns out that such an intrinsic uncertainty
of the fixed-order QCD predictions can be rather large in the
determination of the $W^\pm$ mass: compare the values for $M_{W^\pm}$ 
as obtained in  the NLO fits and
given on the left of Tab.~\ref{tab:CC03_fit} to those on the right. 
The differences between
the reconstructed $M_{W^\pm}$ values for $\mu=35$ and 175 GeV 
 (which, hereafter,
 we denote by $\delta_{\mathrm{NLO}} M_{W^\pm}$) can become as large
as 130 MeV (for the L scheme, in presence of a polynomial background).
The J scheme seems to be here the least sensitive to $\mu$-variations
(a 10 to 20 MeV effect),
with the D and C ones falling in between.  
Notice that, in this exercise,
we have restrained ourselves to values of $\mu$ not smaller than $\sqrt s/5$,
precisely in order to avoid the mentioned logarithmic effects induced
by a choice of $\mu$ too close to the jet-scheme-dependent emission scale 
$\sqrt {y_{{cut}}s}$ (see also Footnote 7). 
Indeed, for $\mu$ in the above interval,
we have found that to change the  values of $y_{\rm{cut}}$ 
(still maintaining the typical four-jet separations used in experimental
analyses) does not affect our conclusions. 
In fact, the latter do not change either, if one adopts a different
recombination procedure of the cluster momenta. 

\section{Conclusions}
\label{sec_summary}

In summary, we have verified that theoretical errors 
arising from the use of different jet clustering algorithms in treating
the fixed-order ${\cal O}(\as)$ corrections from
perturbative QCD, in the prediction
of experimental observables used for the extraction of
the $W^\pm$ boson mass in the hadronic channel $e^+e^-\to
W^+W^-\to4$ jets (CC03) at LEP2, can be competitive with similar
systematic effects that could be induced by non-perturbative dynamics,
such as CR and BECs, e.g., as predicted in the hadronisation
model of Ref.~\cite{EG}. In particular, using various jet definitions, 
and spanning the scale of $\as$ between $M_{W^\pm}$ and
$2M_{W^\pm}$ approximately,
we have shown that these uncertainties on $M_{W^\pm}$ 
can be of order 100 MeV, hence larger than current
experimental assumptions on the size of the theoretical error. 

We also have verified that the inclusion of irreducible background via 
CC11 diagrams has little impact on our main conclusions, so has the 
incorporation of ISR effects. Similarly, a different treatment
of the widths in the resonant propagators (the fixed-width scheme was
adopted here) has negligible consequences for both CC03 and CC11. 
A different choice of $\sqrt s$ (at fixed $y_{\rm{cut}}$'s, or vice versa)
yields similar estimates of $\delta_{\mathrm{NLO}} M_{W^\pm}$ to those 
given here. Also, if one enforces typical $W^\pm$ mass reconstruction cuts,
say, $|M_{R_i}-M_{W^\pm}|<\delta$, for $i=1,2$ and $10$ GeV $<\delta<$
30 GeV, see  eqs.~(\ref{minimise})--(\ref{average}),
typical values of $\delta_{\mathrm{NLO}} M_{W^\pm}$ remain in the above
range, despite the effects on the actual event rates can be dramatic
 \cite{loop}. 

Finally, effects due to the kinematic interplay between jet clustering 
algorithms and PS (including hadronisation) were not in the
original intentions of this study, as they have already been addressed
in Ref.~\cite{schemes}. Whereas the latter can be estimated in the
context of an event level MC analysis, those considered here are intrinsic
uncertainties of the theory. The results of our present analysis
point to the fact that such perturbative QCD effects may not
yet be under control, at least in the context of 
$W^\pm$ mass determinations from the hadronic data
samples collected at LEP2. Taking also into
account the results of Ref.~\cite{schemes} in the same context,
wherein the systematic uncertainties in the
reconstructed $M_{W^\pm}$ value due to the dynamics involved
beyond the hard scattering processes (as obtained in HERWIG and PYTHIA) 
were often found to be somewhat smaller than 100 MeV (
more in line then with the experimental
estimates discussed in the Introduction),
one may conclude that a thorough reassessment of the theoretical
systematics entering the $e^+e^-\to W^+W^-\to$~hadrons channel
is in order, given the importance that the precise knowledge 
of the $W^\pm$ mass has in constraining the properties of yet
undiscovered particles, such as the Higgs boson mass.
This will require the availability of a MC event generator
based on the NLO matrix elements (both real and virtual) used in this analysis
(properly interfaced to the subsequent PS and hadronisation stages),
which is under construction in the HERWIG environment \cite{NLO-MC}.
It will eventually be the implementation of the more sophisticated
selection methods used by the LEP collaborations (as opposed to
the simpler ones illustrated here) in the context of a such a NLO-MC event generator
that will finally assess the uncertainty range on $M_{W^\pm}$ still allowed
by the most up-to-date theoretical and experimental instruments.

\section*{Acknowledgements}

SM is grateful to the Particle Physics Department of
the Rutherford Appleton Laboratory and the Department of Radiation Science
of Uppsala University for the kind hospitality while part of this work was 
carried out. Many useful discussions with Leif L\"onnblad
and Torbj\"orn Sj\"ostrand are also acknowledged by SM.

\end{document}